\documentclass[11pt]{elsart}
\usepackage[dvips]{graphicx}
\usepackage{amsmath}
\usepackage{amssymb}

\begin{document}
\begin{frontmatter}

\title{Counting spanning trees in a small-world Farey graph}

\author[lable1,label2]{Zhongzhi Zhang}
\ead{zhangzz@fudan.edu.cn}
\author[lable1,label2]{Bin Wu}
\author[lable1,label2]{Yuan Lin}

\address[lable1]{School of Computer Science, Fudan
University, Shanghai 200433, China}
\address[label2]{Shanghai Key Lab of Intelligent Information
Processing, Fudan University, Shanghai 200433, China}

\begin{abstract}
The problem of spanning trees is closely related to various
interesting problems in the area of statistical physics, but
determining the number of spanning trees in general networks is
computationally intractable. In this paper, we perform a study on
the enumeration of spanning trees in a specific small-world network
with an exponential distribution of vertex degrees, which is called
a Farey graph since it is associated with the famous Farey sequence.
According to the particular network structure, we provide some
recursive relations governing the Laplacian characteristic
polynomials of a Farey graph and its subgraphs. Then, making use of
these relations obtained here, we derive the exact number of
spanning trees in the Farey graph, as well as an approximate numerical solution for the asymptotic growth constant characterizing the network. Finally, we compare our results
with those of different types of networks previously investigated.
\\

% PACS:

\end{abstract}

%89.20.Hh World Wide Web, Internet
%89.75.Da Systems obeying scaling laws
%89.75.Fb Structures and organization in complex systems
%89.75.-k Complex systems
%89.75.Hc Networks and genealogical trees

\date{\today}
\end{frontmatter}

%%%%%%%%%%%%%%%%%%%%%%%%%%%%%%%%%%%%%%%%%%%%%%%%%%%%%%%%%%%%%%%%%
%%%%%%%%%%%%%%%%%%%%%%%%%%%%%%%%%%%%%%%%%%%%%%%%%%%%%%%%%%%%%%%%%
%\vskip -0.5cm\color{Blue}
%\vbox to 0pt{\kern -14cm {
%\noindent \small \copyright 2005
%{\em Elsevier Science B.V. All rights reserved}\\
%{\em Physica A}, submitted.}
%\vss}\color{Black}

%%%%%%%%%%%%%%%%%%%%%%%%%%%%%%%%%%%%%%%%%%%%%%%%%%%%%%%%%%%%%%%%%%%%
\section{Introduction}

Counting spanning trees in networks (graphs) is a fascinating and
central issue in statistical physics, because of its inherent
relevance to diverse aspects in related fields. For example, the
number of spanning trees is an important measure of reliability of a
network~\cite{Bo86,SzAlKe03}. Again for instance, it is exactly the
number of recurrent configurations of the Abelian sand-pile
models~\cite{Dh90,Dh06,DhMa92}, which have been studied extensively
in the past two decades as a paradigm of the self-organized
criticality~\cite{BaTaWi87,BaTaWi88}. On the other hand, the problem
of spanning trees has numerous connections with other interesting
problems associated with networks, such as dimer
coverings~\cite{TsWu03}, Potts model~\cite{DhMa92,Wu82}, random
walks~\cite{NoRi04,DhDh97}, origin of fractalitity for fractal
scale-free networks~\cite{GoSaKaKi06,KiGoSaOhKaKi07} and many
others~\cite{WuCh04}.

In view of its wide range of applications, the enumeration of
spanning trees has received considerable attentions from the
scientific community. A lot of previous studies have focused on
counting spanning trees on different media, including regular
lattices~\cite{Wu77,ShWu00,Wu02}, the Sierpinski
gaskets~\cite{ChChYa07}, and the Erd\"os-R\'enyi random
graphs~\cite{LyPeSc08}, even scale-free networks~\cite{ZhLiWuZh10,ZhLiWuZo11}
that display the striking power-law degree
distribution~\cite{BaAl99} as found for many real
systems~\cite{AlBa02,DoMe02,Ne03,BoLaMoChHw06}. These researches
uncovered the impacts of various architectures on the number of
spanning trees in different networks. However, not all real-word
networks are scale-free. It has been observed~\cite{AmScBaSt00} that
the degree distribution of some real-life networks (e.g., power
grid) decays exponentially, although they show small-world
effect~\cite{WaSt98} characterized simultaneously by low average
path length and high clustering coefficient. Thus far, the problem
of spanning trees in small-world networks with an exponential degree
distribution has not been addressed.

In this paper, we count spanning trees in a small-world network with
a connectivity distribution decaying exponentially, which is
translated from the Farey sequence~\cite{HaWr79} and thus called
a Farey graph (network). According to the specific structure of the Farey
graph, we derive recurrence formulas for the Laplacian
characteristic polynomials of Farey graph and its subgraphs, based
on which we determine the exact number of spanning trees in Farey
graph and numerical value of its asymptotic growth constant. We also compare our
results with those for other networks with the same average degree
of nodes but different degree distributions.

% and find that it is a little larger
%than 1, which is in comparison with that corresponding to the
%scale-free network~\cite{ZhGaWuZh10,DoGoMe02} with the same average
%degree but without fractality, whose entropy for spanning trees is
%much less than 1. Our research indicates that fractal scaling,
%together with its accompanying properties, has a nontrivial impact
%on the spanning trees in scale-free networks.

\section{Construction and topological properties of Farey graph}

The graph under consideration is derived from the famous Farey
sequence~\cite{HaWr79}. In mathematics, a Farey sequence of order
$n$ ($n$ is a positive integer) is a set (denoted by $F_n$) of
irreducible fractions between 0 and 1 arranged in an increasing
order, the denominators of which do not exceed $n$. For example, the
first four Farey sequences are:  $F_1=\{{0\over 1},{1\over 1}\}$,
$F_2=\{{0\over 1},{1\over 2},{1\over 1}\}$, $F_3=\{{0\over
1},{1\over 3},{1\over 2},{2\over 3},{1\over 1}\}$, $F_4=\{{0\over
1},{1\over 4},{1\over 3},{1\over 2},{2\over 3},{3\over 4},{1\over
1}\}$. In fact, Farey sequence $F_n$ can be constructed from
$F_{n-1}$ by using the Farey sum operation denoted as $\oplus$. Let
$a \over b$ and $c \over d$ be two irreducible fractions, then one
can define their ``mediant'' as $\frac{a}{b} \oplus \frac{c}{d} =
{{a+c} \over {b+d}}$. It has been proved that $F_n$ could be
obtained from $F_{n-1}$ by calculating the mediant between each pair
of consecutive fractions in $F_{n-1}$, keeping only those mediants
with denominator equal to $n$, and placing each mediant between the
two values from which it was derived. The Farey sequence has an
interesting property, that is say, any two neighboring fractions $p
\over q$ and $r \over s$ in a Farey sequence satisfy $rq-ps=1$.

%%%%%%%%%%%%%%%%%%%%%%%%%%%%%%%%%%%%%%%%%%%%%%%%%%%%%%%%%
% Figure  1
%%%%%%%%%%%%%%%%%%%%%%%%%%%%%%%%%%%%%%%%%%%%%%%%%%%%%%%%%%
\begin{figure}[h]
\begin{center}
\includegraphics[width=0.70\linewidth]{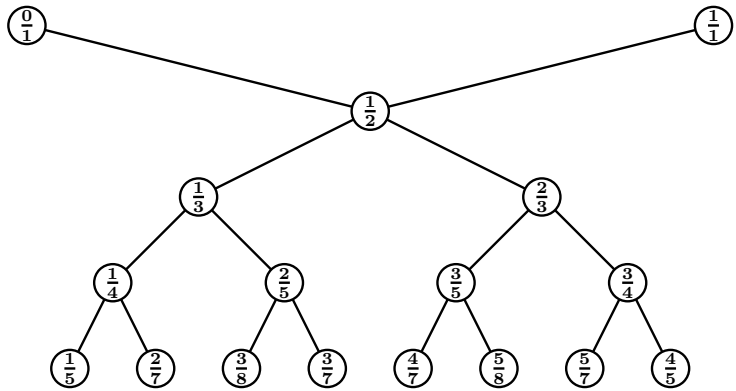}%, angle=90
\end{center}
\caption{Illustration of a Farey tree for the first several
levels.}\label{fig1}
\end{figure}
%%%%%%%%%%%%%%%%%%%%%%%%%%%%%%%%%%%%%%%%%%%%%%%%%%%%%%%%%%

\begin{figure}[h]
\begin{center}
\includegraphics[width=0.80\linewidth]{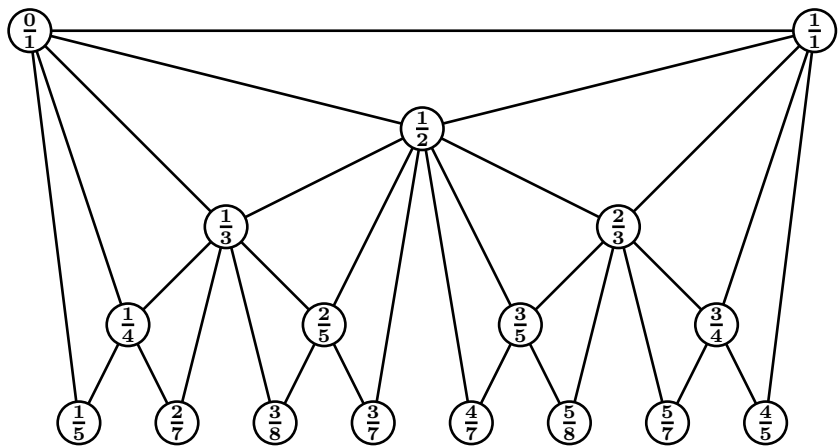}
\end{center}
\caption{Illustration of a Farey graph corresponding to the Farey tree
depicted in Fig. \ref{fig1}.}\label{fig2}
\end{figure}

The Farey sequence can be organized in the so-called Farey
tree~\cite{GoPi85,KiOs86}. Beginning with the two fractions $0 \over
1$ and $1 \over 1$, the first level of the tree is ${1 \over 2}= {0
\over 1} \oplus {1 \over 1}$, and the second level consists of two
fractions $1 \over 3$ and $2 \over 3$ that are obtained by the Farey
sum operation over all the previous fractions, i.e., ${1 \over 3}=
{0 \over 1} \oplus {1 \over 2}$ and ${2 \over 3}= {1 \over 2} \oplus
{1 \over 1}$. Repeating this Farey addition recursively, we can
obtain the Farey tree, the $n$th level of which includes $2^{n-1}$
irreducible fractions, see Fig~\ref{fig1}. In addition, based on the
Farey tree, one can construct a small-world graph called Farey
graph~\cite{Co82,GoPi83}, in which nodes represent the irreducible
fractions between 0 and 1, and two nodes $p \over q$ and $r \over s$
are connected if they satisfy the relation $rq-ps=1$, or
equivalently if they are consecutive terms in some Farey sequence,
see Fig~\ref{fig2}. Notice that, the Farey tree is actually a
spanning tree of the Farey graph.

%%%%%%%%%%%%%%%%%%%%%%%%%%%%%%%%%%%%%%%%%%%%%%%%%%%%%%%%%
% Figure  2
%%%%%%%%%%%%%%%%%%%%%%%%%%%%%%%%%%%%%%%%%%%%%%%%%%%%%%%%%%
%%%%%%%%%%%%%%%%%%%%%%%%%%%%%%%%%%%%%%%%%%%%%%%%%%%%%%%%%%

In fact, the Farey graph can be created in the following iterative
way~\cite{ZhZhWaSh07,ZhCo11}. Let $W_g$ denote the Farey graph after $g$
iterations. For $g=0$, $W_g$ is an edge connecting two nodes. For $g
\geq 1$, $W_{g}$ is obtained from $W_{g-1}$: for each edge in
$W_{g-1}$ introduced at iteration $g-1$, we add a new node and
attach it to both ends of the edge. The Farey graph is minimally
3-colorable, uniquely Hamiltonian, and maximally outerplanar~\cite{Co82,Defintion}. Particularly, the Farey graph exhibits some
remarkable properties of real networks, it is small world with its
average distance increasing logarithmically with its node number,
and its clustering coefficient converges to a large constant $\ln
2$~\cite{ZhZhWaSh07,ZhCo11}.

%%%%%%%%%%%%%%%%%%%%%%%%%%%%%%%%%%%%%%%%%%%%%%%%%%%%%%%%%
% Figure  3
%%%%%%%%%%%%%%%%%%%%%%%%%%%%%%%%%%%%%%%%%%%%%%%%%%%%%%%%%%
\begin{figure}
\begin{center}
\includegraphics[width=.5\linewidth]{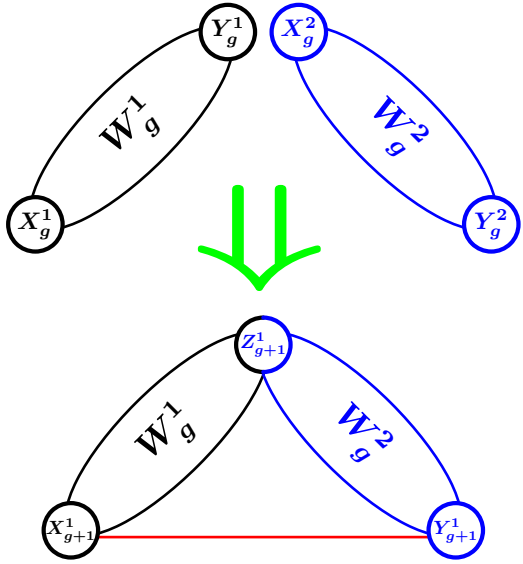}
\caption{(Color online) Alternative construction method of the Farey graph.
$W_{g+1}$ can be obtained by joining two copies of $W_{g}$ denoted
as $W_{g}^{\eta}$ ($\eta=1,2$), the initial nodes of which are
represented by $X_{g}^{\eta}$ and $Y_{g}^{\eta}$, respectively. In
the merging process, $Y_{g}^{1}$ and $X_{g}^{2}$ are identified as
the hub node $Z_{g+1}$ of $W_{g+1}$. In addition, $X_{g}^{1}$ and
$Y_{g}^{2}$ become respectively $X_{g+1}$ and $Y_{g+1}$, which are
linked to each other by a new edge.} \label{merge}
\end{center}
\end{figure}
%%%%%%%%%%%%%%%%%%%%%%%%%%%%%%%%%%%%%%%%%%%%%%%%%%%%%%%%%%

\section{Spanning trees on Farey graph}

After introducing the network construction and its structural
properties, we continue to investigate spanning trees in the Farey
graph $W_g$, by establishing and using recursive relations for the
characteristic polynomials of it and its subgraphs at different
iterations. To facilitate the description, we give the following
definitions. Let $X_g$ and $Y_g$ denote the two nodes in graph $W_g$
that are created at iteration 0 and thus called initial nodes. In
addition, the node in $W_g$ having the highest degree was generated
at iteration 1, it is thus named the hub node and represented by
$Z_g$. Then, the Farey graph can be alternatively built as follows.
Given iteration $g$, $W_{g+1}$ may be obtained by joining at the
initial nodes two copies of $W_{g}$, see Fig.~\ref{merge}. This new
construction shows that the Farey graph is self-similar, implying
that the network order (represented by $N_g$) obeys
$N_g=2N_{g-1}-1$, which coupled with $N_0=2$ gives $N_g=2^g+1$.
Using the self-similar property of the Farey graph, one can
enumerate the number of spanning trees, which is the focus of this
paper.

%\subsection{Characteristic polynomial approach}
%We first introduce the characteristic polynomial approach.

According to the well-known matrix-tree theorem~\cite{Ki1847}, the
number of spanning trees in a connected graph can be expressed
numerically but exactly in terms of the Laplacian spectra
corresponding to the graph. Let $N_{\rm ST}(g)$ represent the number
of spanning trees in the Farey network $W_{g}$, whose Laplacian
matrix is represented by $\textbf{L}_g=[L_{ij}]$, then one can
obtain $N_{\rm ST}(g)$ by computing the product of all non-zero
eigenvalues of $\textbf{L}_g$ as~\cite{Bi93,TzWu00}
\begin{equation}\label{ST01}
N_{\rm ST}(g)=\frac{1}{N_g}\prod_{i=1}^{N_g-1}\lambda_i(g)\,,
\end{equation}
where $\lambda_i(g)$ ($i = 1, 2,\ldots, N_g-1$) denote the $N_g-1$
non-zero eigenvalues of matrix $\textbf{L}_g$, the entry $L_{ij}$ of
which is defined as follows: $L_{ij}=-1$ (or $0$) if nodes $i$ and
$j$ are (or not) adjacent and $i \neq j$, $L_{ij}$ equals the degree
of node $i$ if $i=j$.

Equation~(\ref{ST01}) shows that the issue of determining $N_{\rm
ST}(g)$ is reduced to finding the product of all nonzero Laplacian
eigenvalues. Although the expression of Eq.~(\ref{ST01}) appears
succinct, it requires computing the eigenvalues of a matrix with
order $N_g \times N_g$, which make heavy demands on time and
computational resources for large networks, since the complexity for
calculating eigenvalues is very high. Thus, for large networks, it
is not acceptable to obtain $N_{\rm ST}(g)$ through direct
calculation of the Laplacian spectra, due to the limitations of time
and computer memory. It is then of significant practical importance
to seek for a computationally cheaper approach to overcome this
problem. Fortunately, the self-similar architecture of the Farey
graph allows for calculating the product of its nonzero Laplacian
eigenvalues to obtain an analytical solution to the number of
spanning trees. Details will be provided below.

We use $P_{g}(\lambda)$ to denote the characteristic polynomial of
the Laplacian matrix ${\bf L}_{g}$, i.e.,
\begin{equation}\label{G7}
P_{g}(\lambda)={\rm det}({\bf L}_{g}-\lambda{\bf I}_{g})\,,
\end{equation}
in which ${\bf I}_{g}$ is an $N_g \times N_g$ identity matrix. As
mentioned above, our goal is to evaluate the product of all the
nonzero eigenvalues of ${\bf L}_{g}$, namely, all nonzero roots of
polynomial $P_{g}(\lambda)$.

In order to find the product, we denote ${\bf Q}_{g}$ as an $(N_g-1)
\times (N_g-1)$ submatrix of $({\bf L}_{g}-\lambda{\bf I}_{g})$,
which is obtained by removing from $({\bf L}_{g}-\lambda{\bf
I}_{g})$ the row and column corresponding to an initial node, say,
node $X_g$ of network $W_g$. In addition, we use ${\bf R}_{g}$ to
represent a submatrix of $({\bf L}_{g}-\lambda{\bf I}_{g})$ with an
order $(N_g-2) \times (N_g-2)$, obtained from $({\bf
L}_{g}-\lambda{\bf I}_{g})$ by removing from it the two rows and
columns corresponding to the two initial nodes in $W_g$, i.e., $X_g$
and $Y_g$. On the other hand, let $Q_{g}(\lambda)$ and
$R_{g}(\lambda)$ denote  respectively the determinants of ${\bf
Q}_{g}$ and ${\bf R}_{g}$. Then, the three quantities
$P_{g}(\lambda)$, $Q_{g}(\lambda)$, and $R_{g}(\lambda)$ obey the
following relations:

\begin{equation}\label{G8}
P_{g+1}(\lambda)=\left|\begin{array}{ccc}2d_{g}-\lambda & l_q &
l_q \\ l_q^{\top} & {\bf Q}_{g}+{\bf E}_{g} & -{\bf E}_{g} \\
l_q^{\top} & -{\bf E}_{g} & {\bf Q}_{g}+{\bf
E}_{g}\end{array}\right|,
\end{equation}
\begin{equation}\label{G9}
Q_{g+1}(\lambda)=\left|\begin{array}{ccc}2d_{g}-\lambda & l_q &
l_r \\ l_q^{\top} & {\bf Q}_{g}+{\bf E}_{g} & {\bf O} \\
l_r^{\top} & {\bf O} & {\bf R}_{g}
\end{array}\right|,
\end{equation}
and
\begin{equation}\label{G10}
R_{g+1}(\lambda)=\left|\begin{array}{ccc}2d_{g}-\lambda & l_r & l_r
\\ l_r^{\top} & {\bf R}_{g} & {\bf O} \\ l_r^{\top} & {\bf O} &
{\bf R}_{g}
\end{array}\right|\,.
\end{equation}

In Eqs.~(\ref{G8}-\ref{G10}), ${\bf E}_{g}$ is an $(N_g-1) \times
(N_g-1)$ square matrix with only one entry equal to 1 while all
other entries being 0, viz.,
\begin{eqnarray}\label{L1}
{\bf E}_{g}=\left(\begin{array}{cccc}1 & 0 & \cdots & 0 \\ 0 & 0 &
\cdots & 0 \\ \vdots & \vdots & \quad & \vdots \\ 0 & 0 & \cdots &
0\end{array}\right)\,;
\end{eqnarray}
$2d_{g}$ expresses the degree of the hub node $Z_{g+1}$ in the
$(g+1)$th generation $W_{g+1}$, which is in fact the double of the
degree, $d_{g}$, of an initial node $X_g$ or $Y_g$ in graph $W_{g}$;
$l_{q}$ ($l_{r}$) is a vector of order $N_g-1$ ($N_g-2$) with
$d_{g}$ entries being $-1$ all other $N_g-d_{g}-1$ ($N_g-d_{g}-2$)
entries equaling 0, in which each entry $-1$ describes an edge
connecting the hub node $Z_{g+1}$ and a node belonging to either
$W_{g}^{1}$ or $W_{g}^{2}$, both of which are amalgamated into
$W_{g+1}$; finally, the superscript $\top$ of a vector represents
transpose.

For the convenience of following description, we index respectively
the rows and columns of $P_{g+1}(\lambda)$ (or its variants obtained
from it by elementary matrix operations) by $V_{i}$ and $C_{i}$ with
$i=1,2,\ldots,2N_{g}-1$. Applying the elementary matrix operation,
i.e., replacing column $C_{j} (N_{g}+1\le j \le 2N_{g}-1)$ by
$C_{j}-C_{j-N_{g}+1}$, we have
\begin{eqnarray}\label{L2}
P_{g+1}(\lambda)=\left|\begin{array}{ccc}2d_{g}-\lambda & l_q &
{\bf 0} \\ l_q^{\top} & {\bf Q}_{g}+{\bf E}_{g} & -({\bf Q}_{g}+2{\bf E}_{g}) \\
l_q^{\top} & -{\bf E}_{g} & {\bf Q}_{g}+2{\bf
E}_{g}\end{array}\right|.
\end{eqnarray}
We continue to replace the row $V_{i} (2 \le i \le N_{g})$ in
Eq.~(\ref{L2}) by $V_{i}+V_{i+N_{g}-1}$, yielding
\begin{eqnarray}\label{L3}
P_{g+1}(\lambda)&=&\left|\begin{array}{ccc}2d_{g}-\lambda & l_q &
{\bf 0} \\ 2l_q^{\top} & {\bf Q}_{g} & {\bf O} \\
l_q^{\top} & -{\bf E}_{g} & {\bf Q}_{g}+2{\bf
E}_{g}\end{array}\right|\nonumber\\
&=&\left|\begin{array}{cc}2d_{g}-\lambda & l_q \\ 2l_q^{\top} & {\bf
Q}_{g}\end{array}\right|\left|{\bf Q}_{g}+2{\bf E}_{g}\right|.
\end{eqnarray}
Thus, we have expressed the $P_{g+1}(\lambda)$ as a product of two
determinants, which we denote by $P_{g+1}^{(1)}(\lambda)$ and
$P_{g+1}^{(2)}(\lambda)$, respectively. These two determinants can
be easily evaluated as
\begin{eqnarray}\label{L4}
P_{g+1}^{(1)}(\lambda)&=&\left|\begin{array}{cc}2d_{g}-2\lambda & l_q \\
2l_q^{\top} & {\bf
Q}_{g}\end{array}\right|+\left|\begin{array}{cc}\lambda &
{\bf 0} \\ 2l_q^{\top} & {\bf Q}_{g}\end{array}\right|\nonumber\\
&=&2P_{g}(\lambda)+\lambda Q_{g}(\lambda)
\end{eqnarray}
and
\begin{eqnarray}\label{L5}
P_{g+1}^{(2)}(\lambda)&=&\left|{\bf Q}_{g}\right|+2\left|{\bf
R}_{g}\right|=Q_{g}(\lambda)+2R_{g}(\lambda).
\end{eqnarray}
Inserting Eqs.~(\ref{L4}) and (\ref{L5}) into Eq.~(\ref{L3}), we
have
\begin{eqnarray}\label{L6}
P_{g+1}(\lambda)&=&[2P_{g}(\lambda)+\lambda
Q_{g}(\lambda)][Q_{g}(\lambda)+2R_{g}(\lambda)]\nonumber\\
&=&2P_{g}(\lambda)Q_{g}(\lambda)+4P_{g}(\lambda)R_{g}(\lambda)\nonumber\\
&\quad&+\lambda[(Q_{g}(\lambda))^{2}+2Q_{g}(\lambda)R_{g}(\lambda)]\,.
\end{eqnarray}

Similar to the computation processes for $P_{g+1}^{(1)}(\lambda)$
and $P_{g+1}^{(2)}(\lambda)$, $Q_{g+1}(\lambda)$ and
$R_{g+1}(\lambda)$ can be calculated as shown in Eqs.~(\ref{A1})
and~(\ref{A2}).
\begin{eqnarray}\label{A1}
Q_{g+1}(\lambda)&=&\left|\begin{array}{ccc}d_{g}-\lambda & l_q &
{\bf 0} \\ l_q^{\top} & {\bf Q}_{g}+{\bf E}_{g} & {\bf O} \\
l_r^{\top} & {\bf O} & {\bf
R}_{g}\end{array}\right|+\left|\begin{array}{ccc}d_{g}-\lambda &
{\bf 0} & l_r \\ l_{q}^{\top} & {\bf Q}_{g}+{\bf E}_{g} & {\bf O}
\\ l_r^{\top} & {\bf O} & {\bf
R}_{g}\end{array}\right|
\nonumber\\
&\quad&+\left|\begin{array}{ccc}\lambda & {\bf 0} &
{\bf 0} \\ l_{q}^{\top} & {\bf Q}_{g}+{\bf E}_{g} & {\bf O} \\
l_{r}^{\top} & {\bf O} & {\bf R}_g\end{array}\right|\nonumber\\
&=&R_{g}(\lambda)[P_{g}(\lambda)+Q_{g}(\lambda)]+Q_{g}(\lambda)[Q_{g}(\lambda)+R_{g}(\lambda)]\nonumber\\
&\quad&+\lambda
R_{g}(\lambda)[Q_{g}(\lambda)+R_{g}(\lambda)],
\end{eqnarray}
\begin{eqnarray}\label{A2}
R_{g+1}(\lambda)&=&\left|\begin{array}{ccc}d_{g}-\lambda & l_{r} &
{\bf 0}
\\ l_{r}^{\top} & {\bf R}_{g} & {\bf O} \\ l_{r}^{\top} & {\bf O} &
{\bf R}_{g}
\end{array}\right|+\left|\begin{array}{ccc}d_{g}-\lambda & {\bf 0} &
l_{r} \\ l_{r}^{\top} & {\bf R}_{g} & {\bf O} \\ l_{r}^{\top} & {\bf
O} & {\bf R}_{g}
\end{array}\right|+\left|\begin{array}{ccc}\lambda & {\bf 0} &
{\bf 0} \\ l_{r}^{\top} & {\bf R}_{g} & {\bf O} \\ l_{r}^{\top} &
{\bf O} & {\bf R}_{g}
\end{array}\right|\nonumber\\
&=&Q_{g}(\lambda)R_{g}(\lambda)+Q_{g}(\lambda)R_{g}(\lambda)+\lambda[R_{g}(\lambda)]^{2}.
\end{eqnarray}

%As shown above, $P_{g+1}(\lambda)$, $Q_{g+1}(\lambda)$ and
%$R_{g+1}(\lambda)$ evolve as:
%\begin{eqnarray}\label{G11}
%P_{g+1}(\lambda)=2P_{g}(\lambda)Q_{g}(\lambda)+4P_{g}(\lambda)R_{g}(\lambda)\nonumber\\
%+\lambda[(Q_{g}(\lambda))^{2}+2Q_{g}(\lambda)R_{g}(\lambda)],
%\end{eqnarray}
%\begin{eqnarray}\label{G12}
%Q_{g+1}(\lambda)&=&P_{g}(\lambda)R_{g}(\lambda)+[Q_{g}(\lambda)]^{2}+2Q_{g}(\lambda)R_{g}(\lambda)\nonumber\\
%&\quad&+\lambda[Q_{g}(\lambda)R_{g}(\lambda)+(R_{g}(\lambda))^{2}],
%\end{eqnarray}
%and
%\begin{eqnarray}\label{G13}
%R_{g+1}(\lambda)=2Q_{g}(\lambda)R_{g}(\lambda)+\lambda[R_{g}(\lambda)]^{2}.
%\end{eqnarray}

Having derived the recursive relations for $P_{g}(\lambda)$,
$Q_{g}(\lambda)$, and $R_{g}(\lambda)$, shown in
Eqs.~(\ref{L6}-\ref{A2}), we proceed to compute the product of the
nonzero roots of polynomial $P_{g}(\lambda)$. Since $P_{g}(\lambda)$
has one and only one root equal to zero, say $\lambda_{0}(g)=0$, to
find this product, we define a new polynomial $\bar{P}_{g}(\lambda)$
as
\begin{equation}\label{G14}
\bar{P}_{g}(\lambda)=\frac{1}{\lambda}P_{g}(\lambda)\,.
\end{equation}
Then, it is evident that
\begin{equation}\label{G15}
\prod_{i=1}^{N_{g}-1}\lambda_{i}(g)=\prod_{i=1}^{N_{g}-1}
\bar{\lambda}_{i}(g)\,,
\end{equation}
in which $\bar{\lambda}_{1}(g), \bar{\lambda}_{2}(g), \ldots,
\bar{\lambda}_{N_{g}-1}(g)$ represent the $N_{g}-1$ roots of
polynomial $\bar{P}_{g}(\lambda)$. Thus, the determination of the
product of nonzero eigenvalues of Laplacian matrix ${\bf L}_{g}$ is
equivalent to calculating the product on the right-hand side (rhs)
of Eq.~(\ref{G15}).

To find the product $\prod_{i=1}^{N_{g}-1} \bar{\lambda}_{i}(g)$, we
express polynomial $\bar{P}_{g}(\lambda)$ in the following form,
i.e.,
$\bar{P}_{g}(\lambda)=\sum_{j=0}^{N_{g}-1}\bar{p}_{g}(j)\lambda^{j}$,
in which $\bar{p}_{g}(j)$ is the coefficient of term $\lambda^{j}$
with degree $j$. Since it is obvious that $\bar{p}_{g}(N_{g}-1)=-1$,
we then have
\begin{equation}\label{G16}
\sum_{j=0}^{N_{g}-1}\bar{p}_{g}(j)\lambda^{j}=-\prod_{i=1}^{N_{g}-1}[\lambda-\bar{\lambda}_{i}(g)]\,.
\end{equation}
According to Vieta's formulas, the following relation holds:
\begin{equation}\label{G17}
\prod_{i=1}^{N_{g}-1} \bar{\lambda}_{i}(g)=-\bar{p}_{g}(0)\,.
\end{equation}
Thus, all we need is to determine the constant term $\bar{p}_{g}(0)$
of polynomial $\bar{P}_{g}(\lambda)$.

From Eqs.~(\ref{L6}-\ref{A2}) it is not difficult to derive the
following recursion equations:
\begin{eqnarray}\label{G18}
\bar{P}_{g+1}(\lambda)=2\bar{P}_{g}(\lambda)Q_{g}(\lambda)+4\bar{P}_{g}(\lambda)R_{g}(\lambda)\nonumber\\
+[Q_{g}(\lambda)]^{2}+2Q_{g}(\lambda)R_{g}(\lambda),
\end{eqnarray}
\begin{eqnarray}\label{G19}
Q_{g+1}(\lambda)&=&[Q_{g}(\lambda)]^{2}+2Q_{g}(\lambda)R_{g}(\lambda)+\lambda[\bar{P}_{g}(\lambda)R_{g}(\lambda)\nonumber\\
&\quad&+Q_{g}(\lambda)R_{g}(\lambda)+(R_{g}(\lambda))^{2}],
\end{eqnarray}
and
\begin{eqnarray}\label{G20}
R_{g+1}(\lambda)=2Q_{g}(\lambda)R_{g}(\lambda)+\lambda[R_{g}(\lambda)]^{2}.
\end{eqnarray}

On the basis of above relations, we can find the value for
$\bar{p}_{g}(0)$. To this end, we give some additional variables.
Let $q_{g}(0)$ and $r_{g}(0)$ be the constant terms of
$Q_{g}(\lambda)$ and $R_{g}(\lambda)$, respectively. According to
Eqs.~(\ref{G18}-\ref{G20}), the three quantities $\bar{p}_{g}(0)$,
$q_{g}(0)$ and $r_{g}(0)$ obey the recursive relations:
\begin{equation}\label{G21}
\bar{p}_{g+1}(0)=2\bar{p}_{g}(0)q_{g}(0)+4\bar{p}_{g}(0)r_{g}(0)+[q_{g}(0)]^{2}+2q_{g}(0)r_{g}(0)\,,
\end{equation}
\begin{equation}\label{G22}
q_{g+1}(0)=[q_{g}(0)]^{2}+2q_{g}(0)r_{g}(0)\,,
\end{equation}
and
\begin{equation}\label{G23}
r_{g+1}(0)=2q_{g}(0)r_{g}(0)\,.
\end{equation}

Plugging Eq.~(\ref{G23}) into Eq.~(\ref{G22}) to obtain
\begin{equation}\label{G24}
q_{g+1}(0)=[q_{g}(0)]^{2}+r_{g+1}(0)\,,
\end{equation}
which can be rephrased  as
\begin{equation}\label{G25}
r_{g+1}(0)=q_{g+1}(0)-[q_{g}(0)]^{2}\,.
\end{equation}
Replacing $r_{g}(0)$ in Eq.~(\ref{G23}) by the expression given on
the rhs of Eq.~(\ref{G25}) leads to
\begin{equation}\label{G26}
q_{g+1}(0)=3[q_{g}(0)]^{2}-2q_{g}(0)[q_{g-1}(0)]^{2}.
\end{equation}
Thus, we obtain the recursive relation governing $q_{g+1}(0)$,
$q_{g}(0)$, and $q_{g-1}(0)$, as shown explicitly
in~Eq.~(\ref{G26}).

Solving Eq.~(\ref{G26}) one can arrive at the formula for
$q_{g}(0)$. For this purpose, we introduce an intermediary quantity
$k_{g}$, defined as
\begin{equation}\label{G27}
k_{g}=q_{g}(0)/[q_{g-1}(0)]^{2}\,,
\end{equation}
making use of which Eq.~(\ref{G26}) can be rewritten as
\begin{equation}\label{G28}
k_{g+1}=3-\frac{2}{k_{g}}\,.
\end{equation}
Considering the initial condition $k_{1}=q_{1}(0)/[q_{0}(0)]^{2}=3$,
Eq.~(\ref{G28}) can be solved to yield
\begin{equation}\label{G29}
k_{g}=\frac{2(-1)^{g}-(-2)^{-g}}{(-1)^{g}-(-2)^{-g}}\,.
\end{equation}

With the obtained exact result for $k_{g}$, we can reword
Eq.~(\ref{G27}) as
\begin{equation}\label{G30}
\ln{q_{g}(0)}=2\ln{q_{g-1}(0)}+\ln{k_{g}}\,.
\end{equation}
Using the initial condition $\ln{q_{1}(0)}=\ln{3}$ and the
expression for $k_g$ provided by Eq.~(\ref{G29}), Eq.~(\ref{G30})
can be solved inductively to obtain
\begin{equation}\label{G31}
\ln{q_{g}(0)}=2^{g-1}\sum_{i=0}^{g-1}2^{-i}\ln{\frac{2^{i+2}-1}{2^{i+1}-1}}\,.
\end{equation}
Thus, we have
\begin{eqnarray}\label{G32}
q_{g}(0)&=&\prod_{i=0}^{g-1}\left(\frac{2^{i+2}-1}{2^{i+1}-1}\right)^{2^{g-i-1}}
\nonumber \\
&=&\left(2^{g+1}-1\right)\prod_{i=2}^{g}\left(2^{i}-1\right)^{2^{g-i}}\,.
\end{eqnarray}

After deriving $q_{g}(0)$, we now are in position to calculate
$\bar{p}_{g}(0)$. Notice that Eq.~(\ref{G21}) can be decomposed into
a product of two terms as
\begin{equation}\label{G33}
\bar{p}_{g+1}(0)=[2\bar{p}_{g}(0)+q_{g}(0)][q_{g}(0)+2r_{g}(0)]\,.
\end{equation}
In addition, Eq.~(\ref{G22}) can be rewritten as
\begin{equation}\label{G34}
q_{g+1}(0)=q_{g}(0)[q_{g}(0)+2r_{g}(0)]\,.
\end{equation}
Then, we have
\begin{equation}\label{G35}
\frac{\bar{p}_{g+1}(0)}{q_{g+1}(0)}=2\frac{\bar{p}_{g}(0)}{q_{g}(0)}+1\,,
\end{equation}
which together with $\bar{p}_{1}(0)/q_{1}(0)=-3$ leads to
\begin{equation}\label{G35}
\frac{\bar{p}_{g}(0)}{q_{g}(0)}=-\left(2^{g}+1\right)\,.
\end{equation}
Thus, we can obtain the explicit formula for $\bar{p}_{g}(0)$:
\begin{equation}\label{G36}
\bar{p}_{g}(0)=-\left(2^{g}+1\right)\left(2^{g+1}-1\right)\prod_{i=2}^{g}\left(2^{i}-1\right)^{2^{g-i}}\,.
\end{equation}

Hence, the number of spanning trees in the Farey graph $W_g$ is:
\begin{equation}\label{SG37}
N_{\rm
ST}(g)=-\frac{\bar{p}_{g}(0)}{N_g}=\left(2^{g+1}-1\right)\prod_{i=2}^{g}\left(2^{i}-1\right)^{2^{g-i}}\,.
\end{equation}
Equation~(\ref{SG37}) is our main result, which is exact and holds
for any legal $g$. Note that Eq.~(\ref{SG37}) can also be derived using another approach provided in the appendix. We thank an anonymous referee reminding us of this nice method.

Using the above-obtained result given by Eq.~(\ref{SG37}), one can
determine the asymptotic growth constant of the spanning trees---an
important quantity characterizing network structure--- for the Farey
graph, which is defined as the limiting value~\cite{BuPe93,Ly05}
\begin{equation}\label{Entropy}
E_{W_g}=\lim_{N_g \rightarrow \infty }\frac{\ln N_{\rm
ST}(g)}{N_g}=\lim_{g \rightarrow \infty }\frac{\ln N_{\rm
ST}(g)}{N_g}\,
\end{equation}
that converges to a constant value $0.9458$, a finite number a
little smaller than 1.

The obtained entropy of spanning trees for $W_g$ can be compared to
those previously found for other media with the same average node
degree as the Farey network. In the pseudofractal fractal web, the
entropy is $0.8959$~\cite{ZhLiWuZh10}, a value less than $0.9458$.
For the square lattice and the two-dimensional Sierpinski gasket,
their entropy of spanning trees are $1.16624$~\cite{Wu77} and
$1.0486$~\cite{ChChYa07}, respectively, both of which are greater
than $0.9458$. Therefore, the number of spanning trees in the Farey
graph is larger than that of the pseudofractal fractal web, but is
smaller than that corresponding to the square lattice or the
two-dimensional Sierpinski gasket. The distinctness lies with the
architecture of these networks. Although they have identical average
node degree, they show quite different degree distributions. Thus
they exhibit disparate distribution of Laplacian spectra that have
been shown to display similar distribution as node
degrees~\cite{ChLuVu03,ZhChYe10,DoGoMe02}, which fundamentally
determine the number of spanning trees.

\section{Conclusions}

In this paper, we have investigated the problem of spanning trees in
a Farey graph with the small-world effect and an exponential
degree distribution. On the basis of its structure self-similarity
and a decimation procedure, we derived some recursion relations of
the Laplacian characteristic polynomials for the Farey graph and its
subgraphs at different iterations. We then applied these useful
relations to enumerate spanning trees in the Farey graph and
obtained the exact number of spanning trees, as well as the numerical result of
asymptotic growth constant. An advantage of our technique lies in
the avoidance of laborious computation of Laplacian spectra that is
needed for a generic method for determining spanning trees in
general networks. We also compared our results with those previously
obtained for other networks. Finally, it should be mentioned that there are many interesting questions about the Farey graph for future research, e.g., determining the number of its sub-trees  when only nodes with denominator less than a given value $n$ are kept.

\subsection*{Acknowledgment}

The authors are grateful to the anonymous referees for their valuable comments and suggestions. This work was supported by the National Natural Science Foundation of China under
Grant No. 61074119. % and the Shanghai Leading Academic Discipline Project No.B114.

\appendix

\section{An alternative method for determining the number of spanning trees in the Farey graph}

Here we introduce simply the idea of another method for enumerating spanning spanning trees in graph $W_g$. Suppose that one considers a spring hamiltonian on $W_g$, where all spring constants are equal, and equal to $K$. Then the spring Hamiltonian is
\begin{equation}\label{Appe01}
H=\sum_{i \thicksim j}\frac{K}{2}(x_i-x_j)^2\,,
\end{equation}
in which $x_i$ is the scalar displacement at node $i$ of the graph, and sum extends over all edges $i \thicksim j$ of the graph.

If we calculate the partition function of this graph, by integrating over all $x_i$, except one node is kept at zero displacement, the partition function is easily seen to involve the determinant of the corresponding Laplacian matrix.
Suppose, we integrate over all nodes except the two initial nodes. Then, the restricted partition function can be written in the form
\begin{equation}\label{Appe02}
W_g(x,y)=A_g \exp \left(-K_g(x-y)^2\right)\,,
\end{equation}
where $x$ and $y$ are the displacements at the two initial nodes, and $K_n$ may be called the effective spring constant between them.

The self-similar structure of the Farey graph implies that we can express $A_{g+1}$ and $K_{g+1}$ in terms of $A_{g}$ and $K_{g}$. In fact,
\begin{equation}\label{Appe03}
W_{g+1}(x,y)=\int {\rm d} zW_g(x,z)W_g(z,y)\,,
\end{equation}
where $z$ is the displacement attached to the hub node. From 
Eq.~(\ref{Appe03}), it is easily seen that $K_{g+1} = \frac{1}{2}K_g +1$. This relation is easily solved explicitly to obtain $K_g$, and we can put this solution into the recursion equation
for $A_g$ to get Eq.~(\ref{SG37}) in the main text.

%%%%%%%%%%%%%%%%%%%%%%%%%%%%%%%%%%%%%%%%%%%%%%%%%%%%%%%%%%%%%%%%%
%%%%%%%%%%%%%%%%%%%%%%%%%%%%%%%%%%%%%%%%%%%%%%%%%%%%%%%%%%%%%%%%%


\begin{thebibliography}{10}
\bibitem{Defintion}
A graph is said to be Hamiltonian if it possesses a Hamiltonian cycle. A Hamiltonian cycle, also called a Hamiltonian circuit, Hamilton cycle, or Hamilton circuit, is a graph cycle (i.e., closed loop) through a graph that visits each node exactly once. An undirected graph is an outerplanar graph if it can be drawn in the plane without crossings in such a way that all of the vertices belong to the unbounded face of the drawing. A maximal outerplanar graph is an outerplanar graph that cannot have any additional edges added to it while preserving outerplanarity.
\end{thebibliography}
\end{document}